\documentstyle[prl,twocolumn,epsfig,aps]{revtex}
\def\figscale{
\begin{figure}[hbt]
\noindent
\begin{center}
   \psfig{figure=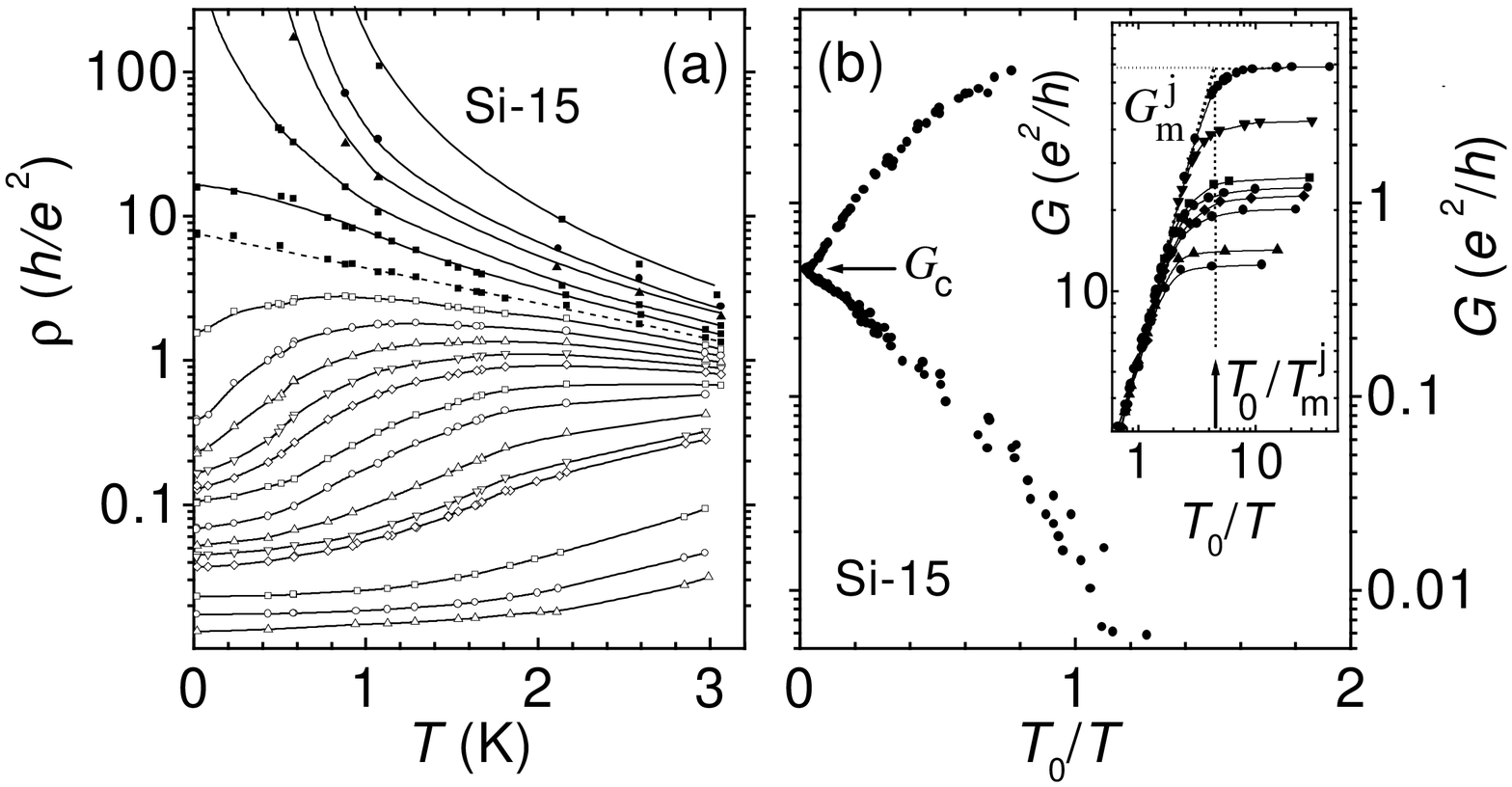,width=0.95\linewidth} 
\end{center}
\caption{(a) Resistivity vs temperature
over the range 20\,mK to 3\,K.
(b) Conductivity vs temperature scaled with a
single parameter $(T_0/T)$.
Different symbols in Fig.~1\,(a) correspond to
the carrier densities
from 0.449 to 0.989 (in steps of 0.054), and further
1.097, 1.205, 1.421, 1.64, 1.745, 2.82, 3.90, $4.985
\times 10^{11}$cm$^{-2}$.
The inset in (b) blows up the metallic part of
the scaling plot in the range of low temperatures.
Different curves correspond to
the densities (from top to bottom)
4.96, 2.81, 1.84, 1.74, 1.63, 1.52, 1.31, and
$1.20 \times 10^{11}$cm$^{-2}$.}
\label{Fig:Scale}
\end{figure}
}

\def\figGc{
\begin{figure}[hbt]
\noindent
\begin{center}
  \psfig{figure=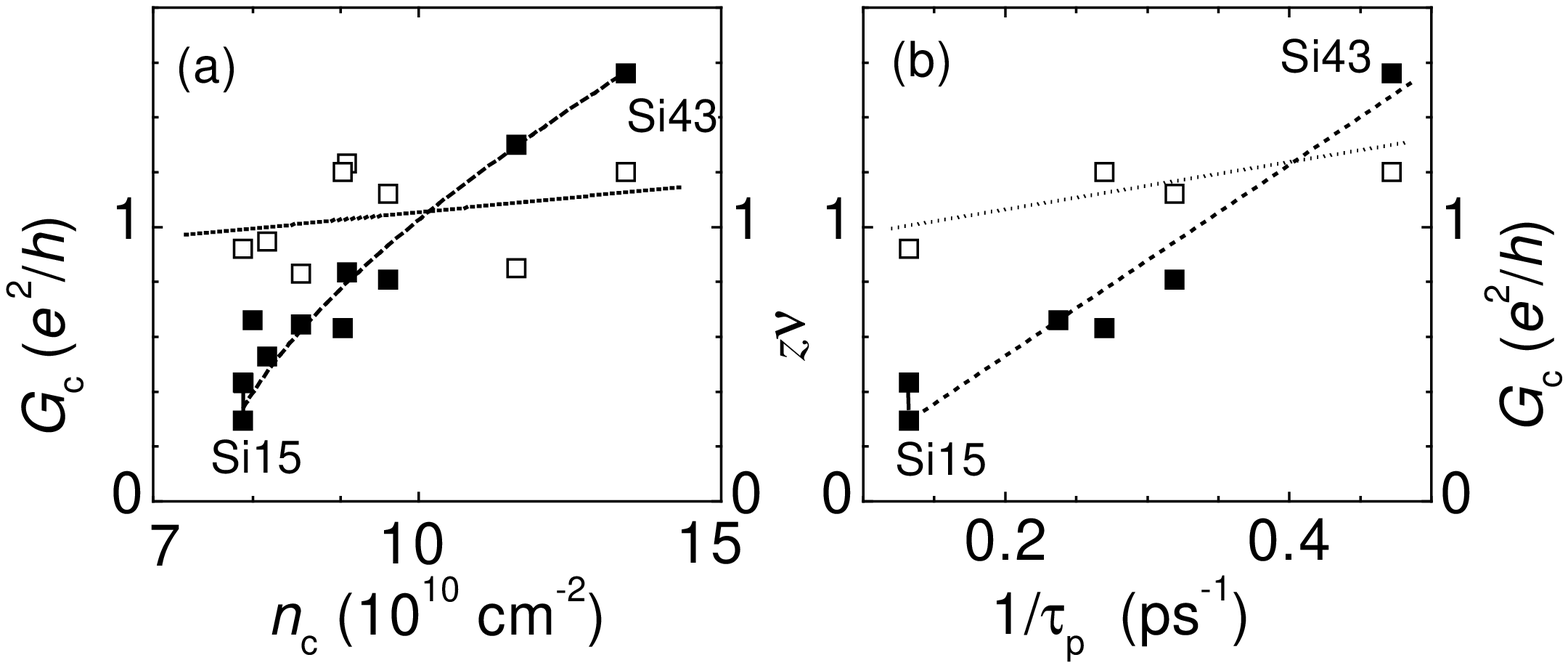,width=0.95\linewidth} 
\end{center}
\caption{Critical conductivity $G_c$
(closed symbols)
and the critical index $z\nu $ (open symbols)
(a) vs critical electron density $n_c$ and
(b) vs elastic scattering time $\tau_{p}$,
for different samples. Dashed and dotted curves
are guides to the eye.}
\label{Fig:Gc}
\end{figure}
}

\def\figlog{
\begin{figure}[hbt]
\noindent
\begin{center}
  \psfig{figure=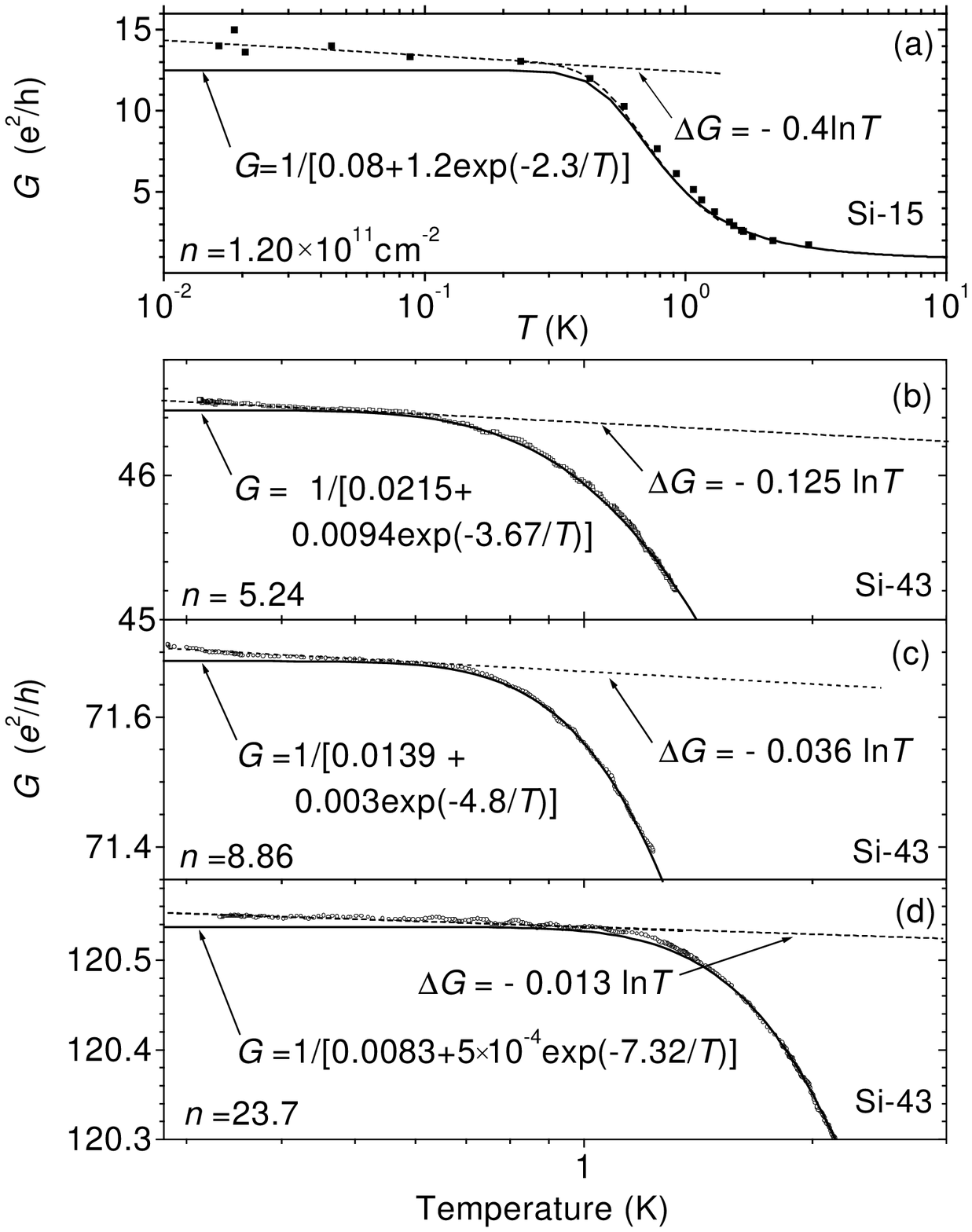,width=0.85\linewidth} 
\end{center}
\caption{Temperature dependence of the conductivity in the low
temperature limit, (a) for the sample Si-15
at low electron density,
and (b), (c) and (d) for the sample Si-43
in a wide range of electron densities
(indicated in units of $10^{11}$ cm$^{-2}$),
respectively. To provide the required signal/noise ratio,
every data point was averaged during 0.5 to 2 minutes.
The continuous lines show the best fits by the exponential
dependence Eq.~(1)
[10],
dashed lines show the logarithmic temperature dependence.}
\label{Fig:log}
\end{figure}
}

\begin{document}
\draft
\title{Logarithmic Temperature Dependence of the Conductivity
and Lack of Universal One-Parameter Scaling in the
Two-Dimensional Metal}
\author{V.~M.~Pudalov$^{(a,b)}$, G.~Brunthaler$^{(b)}$,
A.~Prinz$^{(b)}$, G.~Bauer$^{(b)}$}
\address{$^{(a)} $Institute for High Pressure Physics, Troitsk,
Moscow district 142092, Russia \\
$^({b})$ Institut f\"{u}r Halbleiterphysik,
Johannes Kepler Universit\"{a}t,
Linz, A-4040, Austria}
\date{\today}
\maketitle
\begin{abstract}
We show that the two-dimensional metallic state in Si-MOS samples
persists over a wide range of temperatures (16\,mK to 8\,K), sample
peak mobilities (varying by a factor of 8), carrier densities (0.8 to
$35\times 10^{11}$ cm$^{-2}$) and conductances from 0.3 to
$120 e^2/h$.  Our data reveal a failure of the universal one-parameter
scaling.
We have found a weak delocalizing logarithmic correction
to the conductivity,
which governs {\em a weak increase in the conductivity of the 2D metal
as $T$ approaches zero.}
\end{abstract}

\pacs{PACS numbers: 71.30.+h, 72.15.Rn}

Recent experiments have brought clear evidence
for the occurrence of a metallic state and
a metal-insulator (M-I) transition in
strongly interacting two-dimensional systems \cite{krav94}.
So far such a transition  has been observed  for the two-dimensional
(2D) electron system in Si-MOS structures \cite{krav94,popo97},
for 2D hole and 2D electron systems in Si/SiGe heterostructures
\cite{cole97,ismail97}, and  for the
2D hole system in GaAs/AlGaAs structures \cite{hane97,pepper}.
These findings have initiated a debate over the nature of the 2D metal
\cite{phil97a,phil97b,beli97,puda97a,skvor97,lyan98}
and remain in the focus of experimental and theoretical
interest.

In different materials and samples, the phenomenon is stronger or
weaker, but manifests a number of generic features:\\
(i) The resistivity at zero field drops exponentially as
temperature decreases below
$T_0 \sim 2$\,K \cite{krav94,puda97a}:
    \begin{equation}
      \label{Rexponential}
        \Delta \rho \propto \exp(-(T_0/T)^p),
    \end{equation}
where $p \sim 1$. The drop $\Delta \rho$ is most
pronounced in high mobility structures
\cite{krav94,puda97a}. $T_0$ is sample and density dependent;
for a given sample, it raises
with density  as
$T_0 \propto |\delta n|^{q}$,
where $\delta n = (n-n_c)$ and
$q \approx 1$ \cite{puda97a,hane97}. \\
(ii) The resistivity for each  sample,
can be scaled  into two branches (the metallic
and the insulating ones), using a single parameter
$T/T_0$. The scaling parameter, $T_0$,
thus demonstrates
a critical behavior around the critical density, $n_c$,
\cite{krav94,popo97,cole97,pepper,puda97a},
whereas the exponent $q$ has the meaning of the product,
$q = z\nu$, of the dynamical exponent, $z$,  and of the
correlation length exponent $\nu$ \cite{krav96}.\\
(iii) The metallic state is sensitive to magnetic fields parallel to
the 2D plane \cite{puda97b,simo97,popo97} which points to the
importance of spin effects.\\

These results are in apparent contradiction  to the conventional
picture predicted by  the one-parameter scaling theory (OPST)
for non-interacting particles  \cite{abra79}.
In the experimental systems under study, the interactions are,
no doubt, extremely strong. At the critical density, the ratio
of the Coulomb to Fermi energy,  $E_{ee}/E_F$ is of the order of
10 \cite{puda93}. Besides, due to the broken reflection symmetry in
a triangular potential well, the energy spectrum includes a linear term
$\pm \alpha k_F$. The latter one is large in Si-MOS,
$\alpha \approx 5\times 10^{-6}$\,Kcm,
so that the "energy gap" at the Fermi energy between
the two branches of the spectrum with
different $(\pm)$ chirality,
$\Delta(E_F) = \alpha (k_{F}^{+}+k_{F}^{-}) \geq 5$\,K,
is of the order of the Fermi energy,
$E_F\gtrsim 6$\,K \cite{puda97a,skvor97,revision}.
Thus we believe that the electron system
is spin-polarised in the momentum space, so that the
population ratio of the spin-subbands with different
chirality, $(k_F^{-}/k_F^{+})^2$,
reaches $\approx 2$ for a density of
$n \approx 10^{11}$cm$^{-2}$.
The strong spin polarization may affect the exchange
interaction energy.

Recently, a phenomenologic modyfication of the OPST
has been suggested \cite{dobro97},
to incorporate
also the case of an interacting system,
where the scaling $\beta-$ function
changes the sign at a critical value of the conductance,
$G_c$ in 2D systems.
On the other hand, the theory developed earlier by
Finkel'stein \cite{fink84} 
for systems with strong Coulomb interaction,
predicted a failure of the one parameter scaling picture
and the occurrence of a metallic state.

In this paper we present experimental data which demonstrate the
lack of universality in the one-parameter scaling description of the
conductance.
We studied the high conductivity regime,
$G >G_c \sim 1$ (the conductivity, through the paper,
is in units of  $e^2/h$). We have revealed  the weak
delocalizing logarithmic temperature dependence
of the conductivity persisting down to 16\,mK,
and found
that it gives a positive contribution over the range from
$G \sim 10$ up to $G=120$. Our data
confirm that the 2D metallic state,
once formed by increasing the carrier density $n$ above $n_c$,
remains stable at $n>n_c$.

In order to test  the role of disorder and of interaction,
we have performed systematic measurements on 11 Si-MOS samples
with different disorder.
The samples which were studied most intensively,
had peak mobilities varying by a factor of 8.5:
$\mu=41,000$\,cm$^2$/Vs (Si-15) at $T=0.3$\,K \cite{mobility},
$\mu=40,000$ (Si-5), $\mu=39,000$, $\mu=36,200$ (Si-62),
$\mu=29,000$ (Si-22), $\mu=19,600$ (Si-43),
and $\mu=4,800$\,cm$^2$/Vs (Si-39).
Measurements were taken by a 4-terminal ac-technique
in the temperature range 0.28\,K to 4\,K (for all samples),
and 0.016\,K to 16\,K (for a few samples).
All samples exhibited the characteristic temperature
dependence with a sample dependent critical density
$n_c$, which separates the metallic ($d\rho/dT>0$)
from the insulating regions ($d\rho/dT<0$).
As an example, Fig.~1\,(a) shows
a set of
$\rho(T)-$ curves, taken at different electron densities
for the least disordered sample Si-15.

For samples with lower mobility,
both, the characteristic
magnitude and steepness  of the drop in $\rho(T)$
decrease, whereas the transition shifts to higher densities.
In Si-15, the drop occurs exponentially
(see Eq.\,(1)) \cite{puda97a}
and amounts to $\Delta\rho(T)/\rho(T \ll T_0) \approx 7$,
whereas for Si-39 the drop is almost linear and
within a few percent.
The straight dashed line in Fig.~1\,(a) separates the regions
with positive and negative curvature, $d^2\rho/dT^2 >0$. 
This separatrix line is
almost horizontal for moderate mobility samples (Si-43, Si-22)
and is more and more tilted for less disordered samples as
shown in Fig.~1\,(a) for Si-15. 
For the moderate mobilities, thus, the separatrix
coincides with a ``critical resistivity'', $\rho_c$,
whereas for less disordered samples
the critical resistivity depends
on the studied temperature range and is discussed below.
{\figscale}

The first issue which we wish to explore is the following:
{\em  can the $G(T)$ data be described by
a universal scaling $\beta$-function?}
We note that a non-universal 2D-scaling
for Si-MOS in the insulating state was already discussed earlier
\cite{davies83}.
We explore below the universality of the scaling
function  by examining its major features, namely:
the critical $G_c$-value (at which $\beta(G_c)=0$),
the slope $z\nu$ of the steep part in the vicinity of $G_c$,
and $G_m$ where $\beta(G)$ has it's maximum \cite{dobro97}.

We have performed the scaling analysis on a number of Si-MOS samples.
This procedure is straight forward for moderate mobility samples
with the horizontal $\rho_c(T)$ line. For high mobility samples
where $\rho_c$ is temperature dependent, we limited the scaling
analysis to $T \leq 0.1 E_F/k$.
The scaling plot is shown in Fig.~1\,(b) for this temperature range
for the highest mobility sample.
We obtained the corresponding critical conductance,
the critical density, $n_c$, and the scaling temperature
exponent, $z\nu$,  for each of the samples.
In Fig.~2\,(a) the $G_c$-values for different samples are summarized and
plotted as a function of the critical density $n_c$.
The density $n_c$ is roughly proportional
to the concentration of scatterers at the interface
\cite{puda93}, and thus may serve
as a measure of the disorder.
The momentum relaxation time, $\tau_{p}$, at the maximum mobility for
each of the samples, is another measure of the disorder and thus the variation
of $G_{c}$ is also shown versus $\tau_{p}$ in Figure 2\,(b).
$G_c$ is strongly sample dependent,
raising from $\sim 0.3$ for the least disordered sample Si-15, to
1.8 for sample Si-43.
For sample Si-39 the value of $G_{c} \sim 4$ is not well defined due to the
weak change of $\rho(T)$ in the metallic state and is thus
not shown in Figure 2.
{\figGc}

As is evident from Figs.~2\,(a), and 2\,(b), $G_c(n_c)$ does not
exhibit a saturation at a universal conductance value as disorder
decreases. Instead, these data clearly show the
{\em absence  of such an asymptotic $G_c$ value}.
Thus, at least in the range of
$G_c$  from 0.3 to 1.8 the observed behavior is in contradiction to a
universal behavior.
The inset to Fig.~1\,(b) shows an expanded view of the low
temperature part of $G(T0/T)$ for a few densities $n_j$,
in the metallic range. Each curve $j$ clearly deviates
at low temperatures $T < T_m^j$, from the unified scaling plot.
This deviation was found to be not caused by electron heating, but
is related to the transition of the $\beta(G)$-function
\cite{dobro97} from the steep exponential raise to a gradual decay
\cite{mautern98}. As seen from the inset,
the transition temperature $T_m^j$ (and, hence,
the cut-off length $L_m \propto T_m^{-1/2}$), as well as the
conductance value for the maxima of the $\beta(G)$-function,
$G_m^j$, are not universal.
The absence of a universal scaling behavior already
clearly follows from the tilted $\rho_c(T)$ line in Fig.~1\,(a),
where different $G_c$ values may be calculated by performing
the scaling analysis in different temperature ranges.
The two $G_c$ values for Si-15 shown in Figs.~2
indicate the results obtained from the scaling analysis
over the range of temperatures $T \leq 0.1 E_F$
and over $T \leq 0.15E_F$ (where $E_F \approx 5.8$\,K at $n=n_c$)
\cite{scaling}.

In recent measurements on Si-MOS samples with a backgate, critical
conductivities of $G_{c} = 0.5$ and 0.65 at $n_{c} = 1.67$
and $1.65 \times 10^{11}$\,cm$^{-2}$,
respectively, were reported \cite{popo97}.
Furthermore, two sets of data obtained on
$p-$GaAs/AlGaAs are worth noting:  $G_c=0.65$ and
$n_c= 0.12 \times 10^{11}$\,cm$^{-2}$ \cite{hane97},
and  $G_c=2.1$, $n_c = 0.51\times 10^{11}$\,cm$^{-2}$
\cite{pepper}. For a $n-$Si/SiGe structure,
$G_c=100$ and $n_c=1.8\times 10^{11}$\,cm$^{-2}$,
was reported in Ref.~\cite{ismail97}.
The data taken on different materials or systems with back gate
do not lie on the same dashed line which
interpolates the  $G_c$ vs
$n_c$ data for our Si-MOS samples
and support our conclusion
on the absence of a universal  $G_c$-value.

The onset of the metallic state should occur when the
influence of the disordering (localizing)
potential is compensated by a corresponding  interaction energy.
Figure~2 demonstrates that {\em higher disorder corresponds to higher
$n_c$ and $G_c$ values}.
Since the ratio of Coulomb interaction energy $E_{ee}$ to kinetic
energy $E_F$ decreases with density as
$n^{-1/2}$, we conclude that another interaction mechanism,
whose {\em strength increases with density},
exists apart from the pure Coulomb interaction.
The correlation between
$n_c$, $G_c$ and disorder holds also for the
$p-$GaAs/AlGaAs system, as follows from the
above cited data.

Figure~2 shows also the exponents $z\nu$, obtained from
the scaling analysis on different samples. Although these data are
scattered and less disorder dependent than $G_c$,
there is a clear trend for $z\nu$ to
increase from $z\nu \sim 0.9 \pm 0.1$ for the least disordered
sample, to  $z\nu  \sim 1.6$ for the moderate disordered samples.
This trend is also characteristic for
$p-$GaAs/AlGaAs heterostructures \cite{hane97,pepper}.

For our least disordered sample Si-15, the
transition occurs at a density $n_c  = 0.79\times 10^{11}$cm$^{-2}$
(i.e. about 1.5 times higher than in Ref.~\cite{pepper}),
for the same Coulomb energy
$E_{ee}= (\pi n_s)^{1/2}e^2/\kappa  \approx 70$\,K,
and at a similar Fermi energy ($E_F = 5.6$\,K in Si-15 and 4.6\,K,
in Ref.~\cite{pepper}). The sample mobility in
Ref.~\cite{pepper} is about  25 times
higher, nevertheless, the transition is much weaker
and its features are more similar to those displayed
by our most disordered
sample Si-39 \cite{mautern98},  where the transition occurs at seven times
higher density. This again  confirms our conclusion
on the  involvement of  another  mechanism apart
from the  Coulomb interaction.
We assume that this interaction is caused by spin-effects,
which are particularly enhanced in Si-MOS structures \cite{puda97a}.

As we demonstrated above, no universality either in $G_c$,
in $z\nu $, or in $G_m$ can be found for the Si-MOS samples and
hence, no universal one-parameter $\beta$ function exists.
However, the conductivity
for each sample does display a scaling behavior
in a limited temperature range,
and can thus be described in the framework of a
particular scaling function 
$\beta$.
The occurrence of the M-I transition implies that the
resulting scaling function,  $\beta=d\ln G/d\ln L$,
should change sign at a certain critical
$G_c$-value ($L$ is the system size).
The steep logarithmic part of the $\beta$-function was
plotted in Refs.~\cite{krav95Nott,popo97} and it is
a direct consequence of the exponential temperature
dependence of the resistivity in the metallic phase.
However, the attempts to restore the $\beta$-function
at $G\gg 1$  meet with significant  difficulties
\cite{krav95Nott}. These arise mainly because
the anticipated delocalizing logarithmic corrections to the conductivity
have not been observed until now.
{\figlog}

In order to reveal the weak temperature
correction to the conductance,
measurements have to be performed at large $L_{\rm eff}$,
i.e. at $T \ll T_0$, in the range of the
seeming saturation of the exponential temperature
dependence, Eq.~(1).
This can be achieved most effectively
in the limit of high carrier density
(high $T_0$, and high $G$) and low temperatures.
The measurements in these extreme regimes, are represented in
Fig.~3\,(a) for Si-15 down to the
lowest available temperature of 16~mK at a carrier density of $1.20
\times 10^{11}$ cm$^{-2}$ ($T_0=2.3$\,K)
and in Figs.~3\,(b) to (d) for Si-43
at three different densities from 5.24 to $23.7 \times
10^{11}$ cm$^{-2}$, where $G \sim 120$ and $T_0=7.3$\,K
in the latter case.

Full lines in Figs.~3 represent the empirical
exponential function, $G \sim [\rho_1 + \rho_2 \exp(-T_0/T)]^{-p}$,
where the data were fitted with $p=1$ and
three fitting parameters, $\rho_1$, $\rho_2$, and $T_0$
\cite{puda97a}.
This function saturates as $T\rightarrow 0$, and on this
background one can clearly see a weak temperature
dependence which can be described by
a dependence according to $\Delta G = C\ln T$.
For all samples,
this logarithmic dependence has a negative prefactor, i.e.
it is delocalizing, thus driving the system
to higher conductivity as the temperature decreases.
At relatively low  $G$ (shown in Fig.~3\,(a)),
the prefactor, $C = - 0.4 \pm 0.1$,  is of the order
of the typical
value $u_v/\pi$ \cite{hikami80}, where
$u_v \sim (2 - 1)$ is due the contribution
of two valleys in (100) Si-MOS structures.
At $G=120$, as shown in Fig.~3\,(d), the
prefactor, $C= -0.013 \pm 0.003$,
is about 30 times smaller
than the value quoted above.
At the highest density we studied, $35\times10^{11}$cm$-2$,
the measured prefactor $C= -(6.6 \pm 2) \times 10^{-3} $.

The experimentally observed negative sign
of the logarithmic correction agrees with
recent calculations for a 2D noninteracting system with
broken reflection symmetry \cite{skvor97,lyan98},
whereas the strong reduction of the prefactor $C$ with density
may be related to electron-electron interaction,
ignored in calculations so far.

In summary, we presented experimental evidence for
the absence of universality in the one-parameter scaling
of the conductance of the strongly interacting
2D metallic state in Si-MOS structures.
We have demonstrated the absence
of a second metal-insulator transition and the persistence of a
metallic state in the
conductance range up to $G = 120$, and down to temperatures
as low as 16\,mK. Experimentally, a weak
delocalizing logarithmic correction to the conductivity was found,
which governs {\em the weak increase in the conductivity of the 2D metal
as $T$ approaches 0}.
Furthermore, our  data
reveal the  presence of a  delocalizing
mechanism apart from purely Coulomb interaction
whose strength increases with carrier density.

V.P. acknowledges discussions with M.\ Baranov,
V.\ Kravtsov, Yu.\ Lyanda-Geller, M.\ Skvortsov, I.\ Suslov.
The work was supported by RFBR,
by the Programs on "Physics of solid-state nanostructures" and
"Statistical physics", by INTAS, by NWO the Netherlands,
and by FWF Vienna, ``\"{O}sterreichische Nationalbank'' (project
6333), and GME Austria.


\begin{references}
\bibitem{revision}We have analysed the data for (100)
$n$-Si-MOS structures
and obtained more correct numerical estimate for
the linear term in the energy spectrum:
$\alpha=4.9\times 10^{-6}$\,Kcm, which is 3 times less then
we reported earlier [10].
\bibitem{mobility} We characterise the samples with peak
mobility taken at $T=20$ to 300\,mK, because
in the high $\mu$ samples
the resistivity drops exponentially
at $T<2$\,K. This is in contrast to
the low $\mu$ samples in which  $\rho$ does not change
substantially  at $T < 4$\,K, and which are usually
characterized by mobility measured at 4\,K. Typical
temperature dependence of $\mu$ for our samples
in the range 0.1K to 10\,K is presented in
Ref.~\cite{krav94}.
\bibitem{scaling}It may seem from Fig.~1\,(a) 
that upon further decrease of the considered
temperature range in the scaling analysis,
even lower value, $G_c \sim 0.13$, may be obtained. However, 
$\rho(T)$ data in the $T\rightarrow 0$ limit do not follow
exponential behavior Eq.~(1) and their scaling requires a different
procedure.
\end{references}
\end{document}